\documentclass[conference]{IEEEtran}
\IEEEoverridecommandlockouts
\usepackage{cite}
\usepackage{amsmath,amssymb,amsfonts}
\usepackage{graphicx}
\usepackage{textcomp}
\usepackage{xcolor}
\usepackage{stfloats}

\usepackage{algorithm}
\usepackage{algpseudocode}
\usepackage{amsmath}

\usepackage{multirow}
\usepackage{caption}
\def\BibTeX{{\rm B\kern-.05em{\sc i\kern-.025em b}\kern-.08em
    T\kern-.1667em\lower.7ex\hbox{E}\kern-.125emX}}
\usepackage{geometry}
\begin{document}

\title{Grant-Free Random Access in Uplink LEO Satellite Communications with OFDM\\
%{\footnotesize \textsuperscript{*}Note: Sub-titles are not captured in Xplore and
%should not be used}
\thanks{
The work of Y. Wu is supported in part by the Fundamental Research Funds for the Central Universities, the Yangtze River Delta Science and Technology Innovation Community Joint Research (Basic Research) Project under Grant BK20244006, 111 project BP0719010, and STCSM 22DZ2229005.

R. Mao, Y. Wu, B. Shen, and W. Zhang are with the Department of Electronic Engineering, Shanghai Jiao Tong University, Shanghai 200240, China (e-mails: 
\{maorui2002, yongpeng.wu, boxiao.shen, zhangwenjun\}@sjtu.edu.cn).

S. Chatzinotas is with the Interdisciplinary Center for Security, Reliability and Trust (SnT), University of Luxembourg, 1855 Luxembourg City, Luxembourg and with College of Electronics \& Information, Kyung Hee University, Yongin-si, 17104, Korea (e-mail: Symeon.Chatzinotas@uni.lu).

B. Ottersten is with the Interdisciplinary Center for Security, Reliability and Trust (SnT), University of Luxembourg, 1855 Luxembourg City, Luxembourg (e-mail:bjorn.ottersten@uni.lu).
}
}
\author{Rui Mao, Yongpeng Wu, Boxiao Shen, Symeon Chatzinotas, Bj{\"o}rn Ottersten and Wenjun Zhang} 
\maketitle

\begin{abstract}
This paper investigates joint device activity detection and channel estimation for grant-free random access in Low-earth orbit (LEO) satellite communications. We consider uplink communications from multiple single-antenna terrestrial users to a LEO satellite equipped with a uniform planar array of multiple antennas, where orthogonal frequency division multiplexing (OFDM) modulation is adopted. To combat the severe Doppler shift, a transmission scheme is proposed, where the discrete prolate spheroidal basis expansion model (DPS-BEM) is introduced to reduce the number of unknown channel parameters. Then the vector approximate message passing (VAMP) algorithm is employed to approximate the minimum mean square error estimation of the channel, and the Markov random field is combined to capture the channel sparsity. Meanwhile, the expectation-maximization (EM) approach is integrated to learn the hyperparameters in priors. Finally, active devices are detected by calculating energy of the estimated channel. Simulation results demonstrate that the proposed method outperforms conventional algorithms in terms of activity error rate and channel estimation precision.
\end{abstract}

\begin{IEEEkeywords}
Satellite communications, OFDM, BEM, random access, Doppler shift
\end{IEEEkeywords}

\section{Introduction}
In recent years, the Internet-of-Things (IoT) has gradually entered public awareness and plays an important role in daily life. However, terrestrial cellular communication networks have limited coverage in remote areas and extreme environments, such as oceans, deserts and glaciers\cite{b2}. Low-earth orbit (LEO) satellite with low latency, low power and high flexibility, could provide a promising solution for global coverage\cite{b3}. Therefore, applications of LEO satellite communications in IoT scenarios are attracting considerable attention.

Grant-free random access (GFRA) has been considered as a suitable technique for machine-type communications\cite{b4}, since it allows devices to transmit signals directly and enhances communication efficiency. However, the high mobility of the LEO satellite leads to rapid change in channel state information and large Doppler shifts , resulting in the number of unknown channel parameters much larger than the observations and reducing the performance of existing algorithms. Over the past few years, various methods have been applied in joint device activity detection and channel estimation in satellite communications. For instance, \cite{sbx} utilized channel sparsity in the delay-Doppler-angle domain and proposed a two-dimensional Bayesian learning framework to address the problem. \cite{bit} proposed a training sequences aided modulation architecture and a two-stage successive scheme to refine the system performance. The works mentioned above utilized the orthogonal time frequency space (OTFS) modulation, which has drawbacks in complexity. Additionally, \cite{zaoqi} proposed a Bernoulli–Rician message passing with expectation–maximization (EM) algorithm for LEO-enabled GFRA. \cite{b8} proposed a grid-based parameter probability model and adopted the variance state propagation (VSP) algorithm to leverage sparsity in the delay-Doppler-user domain. However, \cite{zaoqi} considered a slow time-varying channel and \cite{b8} could only handle the residual Doppler shift, after the main delay and Doppler shift were assumed to be precompensated through global navigation satellite system (GNSS).

Basis expansion model (BEM) can approximate the channel model in a subspace, based on a series of orthogonal basis vectors. Therefore, BEM can reduce the number of unknown parameters and is suitable for the high-mobility scenarios\cite{bemearly}. For example, \cite{b10} utilized BEM in end-to-end communications within the LEO satellite scenario and achieved good performance in terms of channel estimation.

In this paper, we investigate joint device activity detection and channel estimation for LEO satellite-enabled GFRA, where multiple single-antenna terrestrial users communicate with a LEO satellite. We assume that only propagation delay is precompensated by the terrestrial devices, while the large Doppler shift and the residual delay are handled at the satellite end. OFDM modulation is adopted to combat the residual delay. To handle the severe Doppler shifts caused by the high mobility of LEO satellite, we propose a transmission scheme, which adopts discrete prolate spheroidal BEM (DPS-BEM) to precisely approximate the channel in the subspace. This largely reduces the number of the unknown channel parameters, since DPS-BEM approximates channel with large Doppler shifts more accurately compared with most of current BEMs\cite{b11}. By introducing the transmission scheme, the joint device activity detection and channel estimation problem can be formulated as a compressive sensing problem. To address the problem, we adopt the vector approximate message passing (VAMP) algorithm\cite{b12}, where two denoisers iterate with each other to estimate the channel. Additionally, Markov random field (MRF) is combined to capture the sparsity in the angular domain. Finally, the EM approach is utilized to update the hyperparameters.

\section{System Model}
We consider a LEO satellite-enabled GFRA system in the uplink transmission, where $U$ single-antenna devices intend to communicate with a LEO satellite. The LEO satellite is equipped with a uniform planar array (UPA) of $N_{y}\times N_{z}$ antennas and a regenerative payload, which is capable of processing baseband signals. The spacing between antennas is half a wavelength. During each time interval, only $U_{a}$ devices utilize the same time-frequency resources to communicate with the satellite. An activity indicator ${\lambda_{u}}$ is introduced for the $u$-th device, where ${\lambda_{u}}$ = 0 if inactive and 1 otherwise. Additionally, following the recommendations of 3GPP\cite{b13}, the propagation delay of each device is precompensated by the terrestrial devices, while the severe Doppler shift and residual delay will be handled at the satellite end. The 
severe Doppler shift is handled by DPS-BEM in our proposed transmission scheme. In this section, we derive the input-output relationship of the communication system and formulate the problem.

\subsection{Input-Output Relationship}
Due to the rapid motion of the LEO satellite and the scatterers around the terrestrial devices, a doubly dispersive channel is considered in the system model. Specifically, each device can potentially transmit $M$ OFDM symbols with $N$ subcarriers. Based on the tap delay line (TDL) model, the received signal of the $(n_{z}+n_{y}N_{z})$-th satellite antenna is expressed as
\begin{equation}\label{tdl}
y^{m}(n,n_{y},n_{z}) = \sum_{u=0}^{U-1}\lambda_{u}\sum_{l=0}^{L-1} h_{u}^{m}(n ; l)x_{u}^{m}(n-l)\varepsilon_{u,n_{y},n_{z}},
\end{equation}
where $n_{y} = 0,...,N_{y}-1$ and $n_{z} = 0,...,N_{z}-1$ are indexes in the space domain; $l = 0,...,L-1$ is index of the tap; $n = 0,...,N-1$, $m = 0,...,M-1$, $u = 0,...,U-1$ and noise is neglected here for simplicity; $x_{u}^{m}(n)$ is the time domain signal modulated from the pilot signal $X_{u}^{m}(k)$, where $k = 0,...,N-1$; $\varepsilon_{u,n_{y},n_{z}} = e^{j\pi n_{z}\mathrm{cos}\varphi_{u}}e^{j\pi n_{y}\mathrm{sin}\varphi_{u}\mathrm{sin}\psi_{u}}$, where $\varphi_{u} \in [0,\pi) $ and $\psi_{u} \in [-\frac{1}{2}\pi,\frac{1}{2}\pi) $  are the azimuth angle and elevation angle of the $u$-th user, respectively. Since LEO satellite operates at an altitude much higher than that of the scatterers around the devices, the angles of each paths associated with the $u$-th device are almost identical\cite{sbx}. The channel response $h_{u}^{m}(n ; l)$ is expressed as\cite{sbx2}
\begin{equation}
    \label{impulse}
h_{u}(n ; l) = \sum\nolimits_{i} a_{i,u}e^{j2\pi\nu_{i,u}nT_{\mathrm{s}}}\mathrm{sinc}(lT_{\mathrm{s}}-\tau_{i,u}),
\end{equation}
where $T_{\mathrm{s}}
$ is the sampling interval. The notations $a_{i,u}$, $\tau_{i,u}$ and $\nu_{i,u}$ are the complex gain, delay and Doppler shift for the $i$-th path from the $u$-th device, respectively. To handle the propagation delay, the cyclic prefix (CP) is introduced, and its duration satisfies $T_{\mathrm{CP}}>L$. The number of taps satisfies $L = \lceil \tau_{\mathrm{max}}/T_{\mathrm{s}}\rceil+1$, where $\tau_{\mathrm{max}}$ is the maximum of delay.

For the convenience of following analysis, we rewrite (\ref{tdl}) as the matrix form. Considering $M$ OFDM symbols transmitted by $U$ potential devices, the input-output relationship in the space domain of the $(n_{z}+n_{y}N_{z})$-th satellite antenna is given by
\begin{equation}\label{ioms}
\mathbf{y}_{n_{y},n_{z}}^{S} = (\mathbf{1}_{M\times UM}\otimes \mathbf{I}_{N}) \mathbf{F} \mathbf{H}_{n_{y},n_{z}}^{S} \mathbf{F}^{H} \mathbf{x} +\boldsymbol{\omega}^{S},
\end{equation}
where $\mathbf{y}_{n_{y},n_{z}}^{S}\in\mathbb{C}^{NM} $ is the received signal; $\boldsymbol{\omega}^{S}$ is the noise; $\mathbf{1}_{M\times UM}$ is an all-ones matrix; $\mathbf{F} = \mathbf{I}_{UM}\otimes \mathbf{F}_{N}$ where $\mathbf{F}^{N}\in\mathbb{C}^{N\times N}$ is the Fourier transform matrix; $\mathbf{x}\in\mathbb{C}^{UNM}$ contains all the pilot signals and the $(mUN+uN+k)$-th element is $X_{u}^{m}(k)$; Space domain channel matrix $\mathbf{H}_{n_{y},n_{z}}^{S}\in\mathbb{C}^{UNM\times UNM}$ is a diagonal block matrix and the $(mU+u)$-th block is $\lambda_{u}\mathbf{H}_{n_{y},n_{z},m,u}^{S}$, where $\mathbf{H}_{n_{y},n_{z},m,u}^{S}\in \mathbb{C}^{N\times N}$ is given by
$\mathbf{H}_{n_{y},n_{z},m,u}^{S}(n,\mathrm{mod}(n-l+N,N)) = {h}_{u}^{m}(n;l)\varepsilon_{u,n_{y},n_{z}}$.

To sufficiently utilize the sparsity in the angular domain, the two-dimensional discrete Fourier transform (2D-DFT) is adopted to the received signal along the space dimension. Then the input-output relationship in the angular domain of the $(a_{z}+a_{y}N_{z})$-th angle is given by

\begin{equation}\label{ioma}
\mathbf{y}_{a_{y},a_{z}}^{A} = (\mathbf{1}_{M\times UM}\otimes \mathbf{I}_{N}) \mathbf{F} \mathbf{H}_{a_{y},a_{z}}^{A} \mathbf{F}^{H} \mathbf{x} +\boldsymbol{\omega}^{A},
\end{equation}
where $a_{y}= 0,...,N_{y}-1$ and $a_{z}=0,...,N_{z}-1$ are indexes in the angular domain; $\boldsymbol{\omega^{A}}$ is the independent zero-mean Gaussian noise with variance $\sigma^{2}$; Angular domain channel matrix $\mathbf{H}_{a_{y},a_{z}}^{A}\in\mathbb{C}^{UNM\times UNM}$ is a diagonal block matrix and the $(mU+u)$-th block $\mathbf{H}_{a_{y},a_{z},m,u}^{A}\in\mathbb{C}^{N\times N}$ is given by
\begin{align}
&\mathbf{H}_{a_{y},a_{z},m,u}^{A}(n,\mathrm{mod}(n-l+N,N)) = \lambda_{u}{h}_{u}^{m}(n;l)\nonumber\\\label{cma}
&\times\Pi_{N_{y}}(a_{y}-\frac{1}{2}N_{y}\mathrm{sin}\varphi_{u}\mathrm{sin}\psi_{u})\Pi_{N_{z}}(a_{z}-\frac{1}{2}N_{z}\mathrm{cos}\varphi_{u}),
\end{align}
where $\Pi_{N}(x) \triangleq \frac{1}{N}\sum_{i=0}^{N-1}e^{-j2\pi\frac{x}{N}i}$. From (\ref{cma}), it is clear that the elements of matrix $\mathbf{H}_{a_{y},a_{z}}^{A}$ have a dominant role only if $a_{y}\approx \frac{1}{2}N_{y}\mathrm{sin}\varphi_{u}\mathrm{sin}\psi_{u}$ and $a_{z}\approx \frac{1}{2}N_{z}\mathrm{cos}\varphi_{u}$, indicating the sparsity in the angular domain.

\subsection{Problem Formulation}
The number of unknown parameters is much greater than that of the observations in (\ref{ioma}), and this poses challenges for joint channel estimation and device activity detection. In this work, a transmission scheme suitable for LEO satellite communications is proposed, where DPS-BEM is adopted to reduce the number of channel parameters. Based on DPS-BEM, channel parameters of the $u$-th device on the $l$-th tap can be expressed as
\begin{equation}\label{bembd}
\mathbf{h}_{l,u} = \sum_{q=0}^{Q-1}g_{q,l}^{u}\mathbf{b}_{q},
\end{equation}
where $\mathbf{h}_{l,u}\in\mathbb{C}^{NM}$ and its $(mN+n)$-th element is $h_{u}^{m}(n;l)$; $g_{q,l}^{u}$ is a projection coefficient on the $q$-th basis vector; $\mathbf{b}_{q}\in\mathbb{C}^{NM}$ is a basis vector, satisfying $\mathbf{b}_{q_{1}}^{H}\mathbf{b}_{q_{1}} = 1$ and $\mathbf{b}_{q_{1}}^{H}\mathbf{b}_{q_{2}} = 0$.

The discrete prolate sequence is used to structure the basis vectors in time domain\cite{b14}. Specifically, $\mathbf{b}_{q}$ is an eigenvector of the matrix $\Theta$,i.e., $\Theta\mathbf{b}_{q} = \lambda_{q}\mathbf{b}_{q}$,
and the matrix is given by
\begin{equation}\label{slpe}
\Theta(a,b) = \frac{\mathrm{sin}[2\pi(a-b)f_{\mathrm{max}}T_{\mathrm{s}}]}{\pi(a-b)},
\end{equation}
where $a,b = 0,1,...,MN-1$, $f_{\mathrm{max}}$ is the maximum of Doppler shift, and $T_{\mathrm{s}}$ is the sampling time. In this work, the selected basis vectors are corresponding to the $Q$ largest eigenvalues of the matrix $\Theta$, and the number of unknown channel parameters is reduced from $ULN_{y}N_{z}MN$ to $ULN_{y}N_{z}Q$. The order $Q$ is usually chosen by $Q \geq \lceil 2Mf_{\mathrm{max}}/\Delta f\rceil+1$. Substituting (\ref{bembd}) into (\ref{ioma}), the input-output relationship is transformed to
\begin{align}
&\mathbf{y}_{a_{y},a_{z}}^{A} 
=(\mathbf{1}_{M\times UM}\otimes \mathbf{I}_{N}) \{ \boldsymbol{\Psi}_{0},...,\boldsymbol{\Psi}_{Q-1} \}\nonumber
\\\label{iombem}
&\times \{\mathbf{I}_{Q}\otimes \mathrm{diag}(\mathbf{x})\mathbf{F}_{L}\}(\mathbf{1}_{M\times 1}\otimes \mathbf{I}_{QLU})\mathbf{g}_{a_{y},a_{z}} + \boldsymbol{\omega}^{A}\\\label{iomgamma}
&= \boldsymbol{\Gamma}  \mathbf{g}_{a_{y},a_{z}} + \boldsymbol{\omega}^{A},
\end{align}
where $\mathbf{F}_{L}=\mathbf{I}_{UM}\otimes\mathbf{F}_{N,L}$ and $\mathbf{F}_{N,L}\in \mathbb{C}^{N\times L}$ is the first $L$ columns of matrix $\mathbf{F}_{N}$; $\boldsymbol{\Psi}_{q} = \mathbf{F}\mathbf{D}_{q}\mathbf{F}^{H}\in \mathbb{C}^{NMU\times NMU}$, where $\mathbf{D}_{q}\in\mathbb{C}^{NMU\times NMU} $ is a diagonal block matrix and the $m$-th block is $\mathbf{I}_{U}\otimes\mathrm{diag}\left\{\mathbf{b}_{q}[mN:(m+1)N-1]\right\}$; The measurement matrix $\boldsymbol{\Gamma}\in\mathbb{C}^{NM\times ULQ}$ consists of known quantities, while the vector $\mathbf{g}_{a_{y},a_{z}}\in\mathbb{C}^{ULQ}$ contains all the unknown parameters and the $(qUL+uL+l)$-th element is expressed as
$g_{q,l}^{u}\lambda_{u}\times\Pi_{N_{y}}(a_{y}-\frac{1}{2}N_{y}\mathrm{sin}\varphi_{u}\mathrm{sin}\psi_{u})\Pi_{N_{z}}(a_{z}-\frac{1}{2}N_{z}\mathrm{cos}\varphi_{u})$.

 Therefore, the joint device activity detection and channel estimation problem is formulated as a sparse signal $\mathbf{g}_{a_{y},a_{z}}$recovery problem. For notation simplicity, we denote the dimension and index of $\mathbf{g}_{a_{y},a_{z}}$ as $E=QLU$ and $e = 0,...,E-1$, respectively, and replace the index $(a_{y},a_{z})$ by $(i,j)$ in the following text.
 
 To perform joint device activity detection and channel estimation, we adopt the Bayesian method, where prior distribution is necessary. Considering the channel sparsity, the Bernoulli-Gaussian (BG) prior distribution is introduced to characterize the parameters\cite{bgem}, i.e.,
\begin{align}
p(g_{i,j}^{e}|s_{i,j}^{e}) 
&= \delta(s_{i,j}^{e}-1) \mathcal{CN}(g_{i,j}^{e}|\mu_{i,j},\phi_{i,j})\nonumber\\\label{pgs}
&+ \delta(s_{i,j}^{e}+1) \delta({g}_{i,j}^{e}),
\end{align}
where $\mu_{i,j}$ and $\phi_{i,j}$ are the mean and variance, respectively, $s_{i,j}^{e}\in \{+1,-1\}$ is the corresponding support. Note that for each channel parameter $g_{i,j}^{e}$ with the same index $(i,j)$, the mean and variance are the same. To precisely describe the sparsity structure of the channel, the MRF is introduced. Then the support can be modelled as\cite{sbx3}
\begin{align}
&p(\mathbf{s}^{e}) 
\propto \mathrm{exp} \left( \sum\limits_{i=0}^{N_{y}-1}\sum\limits_{j=0}^{N_{z}-1}\left( \frac{1}{2}\sum\limits_{s_{i',j'}^{e}\in D_{i,j}} \beta s_{i',j'}^{e}-\alpha \right)s_{i,j}^{e}\right)\nonumber \\\label{pse}
&=\left[\prod_{i,j}\prod_{s_{i',j'}^{e}\in D_{i,j}} \mathrm{exp} (\beta s_{i,j}^{e}s_{i',j'}^{e})\right]^{\frac{1}{2}} \prod_{i,j}\mathrm{exp}(-\alpha s_{i,j}^{e}),
\end{align}
where $\mathbf{s}^{e}$ is a support structure with the $(i,j)$-th element $s_{i,j}^{e}$, the set $D_{i,j}=\left\{ s_{i-1,j}^{e},s_{i+1,j}^{e},s_{i,j-1}^{e},s_{i,j+1}^{e} \right\}$  consists of the neighbors of $s_{i,j}^{e}$, and $\alpha$ and $\beta$ are parameters of the MRF prior. $\beta$ corresponds to the average size of non-zero blocks, and a larger $\alpha$ indicates a sparser $\mathbf{g}$. 

Based on the distribution of $\mathbf{g}_{i,j}$ and the prior noise variance $\sigma^{2}_{i,j}$, the prior distribution of the $(j+iN_{z})$-th angle is given by
\begin{equation}\label{pyg}
p(\mathbf{y}_{i,j},\mathbf{g}_{i,j}) = p(\mathbf{g}_{i,j})\mathcal{CN}(\mathbf{y}_{i,j};\boldsymbol{\Gamma}\mathbf{g}_{i,j},\sigma^{2}_{i,j}\mathbf{I}),
\end{equation}
where $p(\mathbf{g}_{i,j})=p(\mathbf{g}_{i,j}|\mathbf{s}_{i,j})p(\mathbf{s})$ is the prior distribution. In this work, $\mathbf{g}_{i,j}$ is estimated by utilizing the minimum mean square error (MMSE) rule, presented as

\begin{equation}\label{ghat}
\widehat{\mathbf{g}}_{i,j} = \mathrm{arg} \mathop{\mathrm{max}}\limits_{\mathbf{g}'}\mathbb{E}[\|\mathbf{g}'-\mathbf{g}_{i,j}\|_{2}^{2}|\mathbf{y}_{i,j}^{A}].
\end{equation}
Based on that, the channel parameters related to the $m$-th OFDM symbol from the $u$-th device in the angular domain of the $(j+iN_{z})$-th angle can be estimated by
\begin{align}
\widehat{\mathbf{H}}_{i,j,m,u}^{A}&(n,\mathrm{mod}(n-l+N,N))\nonumber\\\label{hhat}
&=\sum\nolimits_{q=0}^{Q-1}{\mathbf{b}}_{q}(mN+n)\widehat{\mathbf{g}}_{i,j}[qUL+uL+l].
\end{align}
Then the active devices can be detected by comparing the energy of the estimated channel with a given threshold
\begin{equation}\label{lmdhat}
\widehat{\lambda}_{u} = \mathbb{I}
\left\{ 
\sum_{i=0}^{N_{y}-1}\sum_{j=0}^{N_{z}-1}\sum_{m=0}^{M-1}\sum_{u=1}^{U} \|\widehat{\mathbf{H}}_{i,j,m,u}^{A}\|_{F}^{2}>\xi
\right\},
\end{equation}
where $\mathbb{I}\{\cdot \}$ is a threshold function, and the threshold $\xi$ is a empirical value to minimize the detection error probability.

\section{Joint Device Activity Detection And Channel Estimation}
In this section, we investigate joint device activity detection and channel estimation by utilizing a compressive sensing algorithm. Firstly, we introduce a factor graph to describe the relationship between variables. Then, the posterior distribution of the variables is estimated by employing the message passing algorithm based on the factor graph. Meanwhile, the EM rule is adopted to update the hyperparameters.

\begin{figure}[h]
    \centering
\includegraphics[width=0.4\textwidth]{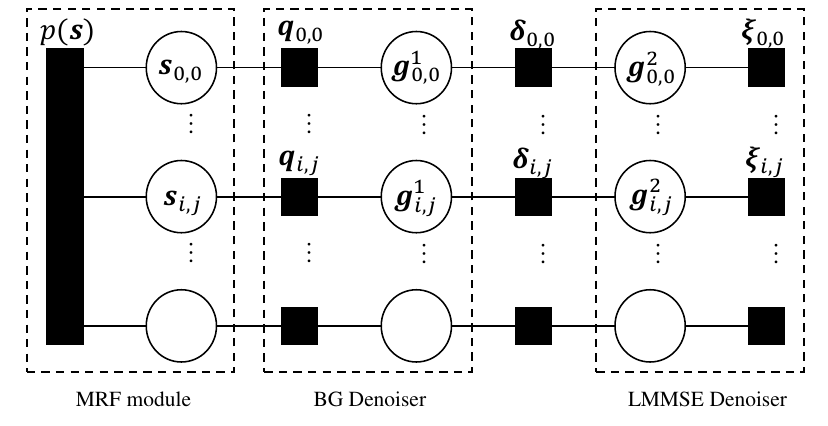}
    \caption{Factor graph}
    \label{fig1}
\end{figure}

\subsection{Factor Graph}
We start with the prior distribution in (\ref{pyg}), and divide $\mathbf{g}_{i,j}$ into two identical variables $\mathbf{g}^{1}_{i,j}=\mathbf{g}^{2}_{i,j}$. The equivalent expression of (\ref{pyg}) is given by
\begin{equation}\label{pyg1g2}
p(\mathbf{y}_{i,j},\mathbf{g}^{1}_{i,j},\mathbf{g}^{2}_{i,j}) = p(\mathbf{g}^{1}_{i,j})\delta(\mathbf{g}^{1}_{i,j}-\mathbf{g}^{2}_{i,j})\mathcal{CN}(\mathbf{y}_{i,j};\boldsymbol{\Gamma}\mathbf{g}^{2}_{i,j},\sigma^{2}_{i,j}\mathbf{I}),
\end{equation}
where $\delta(\cdot)$ is the Dirac distribution. The related factor graph is shown in Fig. 1, which consists of two kinds of nodes:
\begin{itemize}
\item Variable nodes $\{\mathbf{s}_{i,j}\}$, $\{\mathbf{g}^{1}_{i,j}\}$, $\{\mathbf{g}^{2}_{i,j}\}$ are depicted as white nodes, corresponding to the variables with the same names in (\ref{pse}) and (\ref{pyg1g2}).
\item Factor nodes $\{ \boldsymbol{\delta}_{i,j}\}$, $\{\boldsymbol{q}_{i,j}\}$, $\{\boldsymbol{\xi}_{i,j}\}$ are depicted as black nodes, corresponding to the Dirac distribution $\delta(\mathbf{g}^{1}_{i,j}-\mathbf{g}^{2}_{i,j})$, the conditional distribution $p(\mathbf{g}^{1}_{i,j}|\mathbf{s}_{i,j})$ in (\ref{pgs}), and the distribution $\mathcal{CN}(\mathbf{y}_{i,j};\boldsymbol{\Gamma}\mathbf{g}^{2}_{i,j},\sigma^{2}_{i,j}\mathbf{I})$ in (\ref{pyg}), respectively.
\end{itemize}
Additionally, as shown in Fig. 1, the structure is divided into three modules: BG denoiser is a non-linear denoiser aiming to estimate $\mathbf{g}^{1}_{i,j}$, based on the prior distribution of $\mathbf{g}_{i,j}$. Linear minimum mean square error (LMMSE) denoiser is a linear denoiser aiming to estimate $\mathbf{g}^{2}_{i,j}$, based on the prior distribution $\mathcal{CN}(\mathbf{y}_{i,j};\boldsymbol{\Gamma}\mathbf{g}^{2}_{i,j},\sigma^{2}_{i,j}\mathbf{I})$. MRF module captures the sparsity in the angular domain and updates the posterior probability of the channel. During once iteration, BG denoiser and LMMSE denoiser handle the variables with the same index $(i,j)$, while MRF module handles the variables with the same index $e$.

\subsection{Posterior Distribution Estimation}
In this subsection,  based on the known hyperparameters, we describe how the messages pass between the nodes during once iteration. Additionally, we denote $k$, $\varpi_{a}^{b}$, $x^{e}$ as the number of iterations, the message passed from node $a$ to node $b$, and the $e$-th element of vector $\mathbf{x}$, respectively.

We start from the output messages of the MRF module, which contain the probability that variable $\mathbf{g}_{i,j}^{1}$ equals to zero. The message from $q_{i,j}^{e}$ to $g_{i,j}^{1,e}$ is a BG distribution given by
\begin{equation}\label{xxqg}
\begin{aligned}
&\varpi_{q_{i,j}^{e}}^{g_{i,j}^{1,e}} 
 \propto \int_{s_{i,j}^{e}}p(g_{i,j}^{1,e}|s_{i,j}^{e})\varpi_{s_{i,j}^{e}}^{q_{i,j}^{e}}\\
 &= \pi_{s_{i,j}^{e}\rightarrow q_{i,j}^{e}} \mathcal{CN}(g_{i,j}^{1,e}|\mu_{i,j},\phi_{i,j})+(1-\pi_{s_{i,j}^{e}\rightarrow q_{i,j}^{e}}) \delta({g}_{i,j}^{1,e}),
\end{aligned}
\end{equation}
where $\pi_{s_{i,j}^{e}\rightarrow q_{i,j}^{e}}$ is the posterior probability updated by MRF module. Attributed to the i.i.d. assumption, the sum-product (SP) belief of $\mathbf{g}_{i,j}^{1}$ can be independently computed by
\begin{align}
&b_{\mathrm{sp}}(g^{1,e}_{i,j}) \propto p(g^{1,e}_{i,j}|\mu_{i,j},\phi_{i,j}) \mathcal{CN}(g^{1,e}_{i,j};{r}^{1k,e}_{i,j}, (\gamma^{1k}_{i,j})^{-1})\nonumber \\\label{bspg}  
&=\pi_{i,j}^{k,e}\mathcal{CN}(g_{i,j}^{1,e};\mu_{i,j}^{k,e},\phi_{i,j}^{k,e})+(1-\pi_{i,j}^{k,e})\delta(g_{i,j}^{1,e}),
\end{align}
where
\begin{subequations}
\begin{align}\label{pikeij}
\pi_{i,j}^{k,e} =& \left( 1+ \frac{(1-\pi_{s_{i,j}^{e}\rightarrow q_{i,j}^{e}})\mathcal{CN}\left(0;r_{i,j}^{1k,e},(\gamma_{i,j}^{1k})^{-1}\right)}{\pi_{s_{i,j}^{e}\rightarrow q_{i,j}^{e}}\mathcal{CN}\left(0;r_{i,j}^{1k,e}-\mu_{i,j},\frac{1}{\gamma_{i,j}^{1k}}+\phi_{i,j}\right)}\right)^{-1},\\\label{miukeij}
\mu_{i,j}^{k,e} =& \frac{\gamma_{i,j}^{1k}r_{i,j}^{1k,e}+\mu_{i,j}(\phi_{i,j})^{-1}}{\gamma_{i,j}^{1k}+(\phi_{i,j})^{-1}},\\\label{phikeij}
\phi_{i,j}^{k,e} =&\left( \gamma_{i,j}^{1k}+(\phi_{i,j})^{-1}\right)^{-1}.
\end{align}
\end{subequations}
To describe the performance of the BG denoiser, a condition-mean estimator $f_{1}(r^{1k,e}_{i,j},\gamma^{1k}_{i,j}) = \mathbb{E}\left[ g^{1,e}_{i,j}|r^{1k,e}_{i,j},\gamma^{1k}_{i,j}\right]$ is introduced, then $\mathbf{g}_{i,j}^{1k}$ can be approximated by computing the expectation of its SP belief, given by
\begin{align}
\widehat{\mathbf{g}}_{i,j}^{1k} 
= \mathbf{f}_{1}(\mathbf{r}_{i,j}^{1k},\gamma_{i,j}^{1k})
=\boldsymbol{\pi}_{i,j}^{k}\odot \boldsymbol{\mu}_{i,j}^{k},\label{ghat1ij}
\end{align}
where $\boldsymbol{\pi}_{i,j}^{k}=\left[\pi_{i,j}^{k,0},...,\pi_{i,j}^{k,E-1} \right]^{T}$, $\boldsymbol{\mu}_{i,j}^{k}=\left[\mu_{i,j}^{k,0},...,\mu_{i,j}^{k,E-1} \right]^{T}$, and the approximate noise precision can be computed by $\eta^{1k}_{i,j} = \left(\gamma^{1k}_{i,j} +(\phi_{i,j})^{-1} \right)/\left\langle \boldsymbol{\pi}_{i,j}^{k}\right\rangle$, where $\left\langle\mathbf{x}\right\rangle$ is the average of vector $\mathbf{x}$. Since the approximate belief of $\mathbf{g}_{i,j}^{1}$ is given by $b_{\mathrm{app}}(\mathbf{g}^{1}_{i,j}) = \mathcal{CN}(\mathbf{g}^{1}_{i,j}; \widehat{\mathbf{g}}^{1k}_{i,j}, (\eta^{1k}_{i,j})^{-1}\mathbf{I})$, according to message passing rules, $\varpi_{\mathbf{g}_{i,j}^{1}}^{\boldsymbol{\delta}_{i,j}}$ is a Gaussian distribution given by
\begin{equation}\label{xxg1delta}
\varpi_{\mathbf{g}_{i,j}^{1}}^{\boldsymbol{\delta}_{i,j}} 
= \mathcal{CN}(\mathbf{g}^{1}_{i,j};\mathbf{r}^{2k}_{i,j},(\gamma^{2k}_{i,j})^{-1}\mathbf{I}),   
\end{equation}
where $\mathbf{r}^{2k}_{i,j} = (\eta^{1k}_{i,j}\widehat{\mathbf{g}}^{1k}_{i,j}-\gamma^{1k}_{i,j}\mathbf{r}^{1k}_{i,j})/(\eta^{1k}_{i,j}-\gamma^{1k}_{i,j})$ and $\gamma^{2k}_{i,j} = \eta^{1k}_{i,j}-\gamma^{1k}_{i,j}$ are the mean and variance, respectively.

The processes of LMMSE denoiser are similar. The approximate belief of $\mathbf{g}_{i,j}^{2}$ is $b_{\mathrm{app}}(\mathbf{g}^{2}_{i,j}) = \mathcal{CN}(\mathbf{g}^{2}_{i,j}; \widehat{\mathbf{g}}^{2k}_{i,j}, (\eta^{2k}_{i,j})^{-1}\mathbf{I})$ and $(\eta^{2k}_{i,j})^{-1}$ is the approximate noise variance. Similarly, to describe the performance of the LMMSE denoiser, a condition-mean estimator $f_{2}(r^{2k,e}_{i,j},\gamma^{2k}_{i,j}) = \mathbb{E}\left[ g^{2,e}_{i,j}|r^{2k,e}_{i,j},\gamma^{2k}_{i,j}\right]$ is introduced, then the approximation of $\mathbf{g}_{i,j}^{2k}$ is given by
\begin{align}
\widehat{\mathbf{g}}^{2k}_{i,j} 
&= \mathbf{f}_{2}(\mathbf{r}^{2k}_{i,j},\gamma^{2k}_{i,j})\nonumber\\\label{ghat2ij}
&= \left( \sigma_{i,j}^{-2} \boldsymbol{\Gamma}^{H}\boldsymbol{\Gamma} + \gamma_{i,j}^{2k}\mathbf{I}
\right)^{-1}
\left( \sigma_{i,j}^{-2} \boldsymbol{\Gamma}^{H}\mathbf{y}_{i,j} + \gamma_{i,j}^{2k}\mathbf{r}_{i,j}^{2k}
\right),
\end{align}
and the approximate noise precision can be computed by $\eta^{2k}_{i,j} = MN/\sum_{n}\frac{1}{\gamma^{2k}_{i,j}+|s_{n}|^{2}\sigma_{i,j}^{-2}}$, where $s_{n}$ is a singular value of matrix $\boldsymbol{\Gamma}$. Since the approximate belief of $\mathbf{g}_{i,j}^{2}$ is given by $b_{\mathrm{app}}(\mathbf{g}^{2}_{i,j}) = \mathcal{CN}(\mathbf{g}^{2}_{i,j}; \widehat{\mathbf{g}}^{2k}_{i,j}, (\eta^{2k}_{i,j})^{-1}\mathbf{I})$, the message from $\mathbf{g}_{i,j}^{2}$ to $\boldsymbol{\delta}_{i,j}$ is a Gaussian distribution given by
\begin{equation}\label{xxg2delta}
\varpi_{\mathbf{g}_{i,j}^{2}}^{\boldsymbol{\delta}_{i,j}} 
= \mathcal{CN}(\mathbf{g}_{i,j}^{2};\mathbf{r}_{i,j}^{1,k+1},(\gamma_{i,j}^{1,k+1})^{-1}\mathbf{I}),
\end{equation}
where $\mathbf{r}_{i,j}^{1,k+1} = (\eta_{i,j}^{2k}\widehat{\mathbf{g}}_{i,j}^{2k}-\gamma_{i,j}^{2k}\mathbf{r}_{i,j}^{2k})/(\eta_{i,j}^{2k}-\gamma_{i,j}^{2k})$ and $\gamma_{i,j}^{1,k+1} = \eta_{i,j}^{2k}-\gamma_{i,j}^{2k}$ are the mean and variance, respectively.

In the MRF module, messages pass between the nodes $\{ \mathbf{s}^{e} \}$ with the same index $e$. We start from the input messages of the MRF module, with $\varpi_{\boldsymbol{\delta_{i,j}}}^{\mathbf{g}_{i,j}^{1}}=\mathcal{CN}(\mathbf{g}_{i,j}^{1};\mathbf{r}_{i,j}^{1,k+1},(\gamma_{i,j}^{1,k+1})^{-1}\mathbf{I})$, the message from $q_{i,j}^{e}$ to $s_{i,j}^{e}$ is a Bernoulli distribution given by
\begin{align}
&\varpi_{q_{i,j}^{e}}^{s_{i,j}^{e}}
\propto \int_{g_{i,j}^{1,e}}p(g_{i,j}^{1,e}|s_{i,j}^{e})\varpi_{\delta_{i,j}^{e}}^{g_{i,j}^{1,e}}\nonumber\\\label{xxqs}
&=\pi_{q_{i,j}^{e}\rightarrow s_{i,j}^{e}}\delta(s_{i,j}^{e}-1)+(1-\pi_{q_{i,j}^{e}\rightarrow s_{i,j}^{e}})\delta(s_{i,j}^{e}+1),
\end{align}
where
\begin{equation}\label{piqs}
\pi_{q_{i,j}^{e}\rightarrow s_{i,j}^{e}} = \left( 1+ \frac{\mathcal{CN}\left(0;r_{i,j}^{2k,e},(\gamma_{i,j}^{2k})^{-1}\right)}{\mathcal{CN}\left(0;r_{i,j}^{2k,e}-\mu_{i,j},(\gamma_{i,j}^{2k})^{-1}+\phi_{i,j}\right)}\right)^{-1}.
\end{equation}

To adapt to the UPA mentioned above, the two-dimensional MRF is adopted, where the nodes are 4-connected. To succinctly describe the relative position, we denote the top, bottom, left, right neighbors to $s_{i,j}^{e}$ as $D_{i,j} = \left\{ s_{i,jt}^{e},s_{i,jb}^{e},s_{i,jl}^{e},s_{i,jr}^{e}  \right\}$, (i.e., $s_{i,jt}^{e} = s_{i,j+1}^{e}$, $s_{i,jb}^{e} = s_{i,j-1}^{e}$, $s_{i,jl}^{e} = s_{i-1,j}^{e}$, $s_{i,jr}^{e} = s_{i+1,j}^{e}$) and denote the messages from top, bottom, left, right to $s_{i,j}^{e}$ as $\varpi_{i,j}^{e,t}$, $\varpi_{i,j}^{e,b}$, $\varpi_{i,j}^{e,l}$, $\varpi_{i,j}^{e,r}$, respectively. Then the input message from left of node $s_{i,j}^{e}$ is given by
\begin{equation}\label{xxleft}
\varpi_{i,j}^{e,l} = \pi_{i,j}^{e,l}\delta\left( s_{i,j}^{e}-1 \right) + \left(1-\pi_{i,j}^{e,l}\right)\delta\left( s_{i,j}^{e}+1 \right),
\end{equation}
where $\pi_{i,j}^{e,l}$ is given by (27) at the bottom of next page, and the expressions of $\varpi_{i,j}^{e,b}$, $\varpi_{i,j}^{e,t}$, $\varpi_{i,j}^{e,r}$ are similar. Then, the message from $s_{i,j}^{e}$ to $q_{i,j}^{e}$ is a Bernoulli distribution given by
\begin{figure*}[hb]
\hrulefill
\begin{align}
\pi_{i,j}^{e,l} = \frac
{\pi_{q_{i,jl}^{e}\rightarrow s_{i,jl}^{e}}\prod_{d\in\{l,t,b\}}\pi_{i,jl}^{e,d}e^{-\alpha+\beta}+\left(1-\pi_{q_{i,jl}^{e}\rightarrow s_{i,jl}^{e}}\right) \prod_{d\in\{l,t,b\}}\left( 1-\pi_{i,jl}^{e,d} \right)e^{\alpha-\beta}}
{\left( e^{\beta} + e^{-\beta} \right) \left(\pi_{q_{i,jl}^{e}\rightarrow s_{i,jl}^{e}}e^{-\alpha}\prod_{d\in\{l,t,b\}}\pi_{i,jl}^{e,d} + (1-\pi_{q_{i,jl}^{e}\rightarrow s_{i,jl}^{e}})e^{\alpha} \prod_{d\in\{l,t,b\}}( 1-\pi_{i,jl}^{e,d} )\right)} 
\end{align}
\end{figure*}
\begin{equation}\label{xxsq}
\varpi_{s_{i,j}^{e}}^{q_{i,j}^{e}}=\pi_{s_{i,j}^{e}\rightarrow q_{i,j}^{e}}\delta (s_{i,j}^{e}-1)+(1-\pi_{s_{i,j}^{e}\rightarrow q_{i,j}^{e}})\delta (s_{i,j}^{e}+1),
\end{equation}
where
\begin{equation}\label{pisq}
\pi_{s_{i,j}^{e}\rightarrow q_{i,j}^{e}} = \frac
{e^{-\alpha}\prod_{d\in\{t,b,l,r\}}\pi_{i,j}^{e,d}}
{e^{-\alpha}\prod_{d\in\{t,b,l,r\}}\pi_{i,j}^{e,d}+e^{\alpha}\prod_{d\in\{t,b,l,r\}}(1-\pi_{i,j}^{e,d})}
\end{equation}

During once iteration, the messages traverse all the nodes. Two denoisers pass messages via node $\boldsymbol{\delta}$ to reduce the noise and MRF module captures the sparsity in the angular domain.

\subsection{Learning the Hyperparameters}
In this subsection, the EM algorithm is adopted to update the hyperparameters, the prior of which are denoted as $\vartheta \triangleq \left\{
\sigma_{i,j}^{2}, \mu_{i,j}, \phi_{i,j}
\right\}
$. In the $k$-th iteration, the hyperparameters are updated by
\begin{equation}\label{emzong}
\vartheta^{k+1} =  \mathrm{arg} \mathop{\mathrm{max}}\limits_{\vartheta}\mathbb{E}\left[\mathrm{log}\thinspace p(\mathbf{g}_{i,j},\mathbf{y}_{i,j}|\vartheta)|\mathbf{y}_{i,j},\vartheta^{k}   \right].
\end{equation}
By calculating (\ref{emzong}), the closed-form solutions are given by
\begin{subequations}\label{emsmp}
    \begin{align}\label{emsigma}
        \left(\sigma^{2}_{i,j}\right)^{k+1} &=  
        \frac{1}{E} \|
        \mathbf{y}_{i,j}-\boldsymbol{\Gamma}\mathbf{r}_{i,j}^{2k}
        \|^{2}
        +\frac{1}{E}\sum_{n}\frac
        {|s_{n}|^{2}}
        {|s_{n}|^{2}/(\sigma^{2}_{i,j})^{k}+\gamma_{i,j}^{2,k}},
        \\\label{emmiu}
        \left(\mu_{i,j}\right)^{k+1} &=\frac{\sum\nolimits_{e}\pi_{i,j}^{k,e}\mu_{i,j}^{k,e}}{\sum\nolimits_{e}\pi_{i,j}^{k,e}},
        \\\label{emphi}
\left(\phi_{i,j}\right)^{k+1} &= \frac{\sum\nolimits_{e}\pi_{i,j}^{k,e}(|\mu_{i,j}^{k,e}-(\mu_{i,j})^{k}|^{2}+\phi_{i,j}^{k,e})}{\sum\nolimits_{e}\pi_{i,j}^{k,e}}.
    \end{align}
\end{subequations}
The EM-MRF-VAMP algorithm is summarized in Algorithm 1. In this work, we adopt the SVD form of the VAMP algorithm to reduce the complexity. The complexity of the proposed algorithm mainly depends on the VAMP algorithm and the MRF module, and the total complexity is $\mathcal{O}(N_{y}N_{z}QLUMN+N_{y}N_{z}QLUK_{\mathrm{mrf}})$, which is linear with the number of users and is suitable for GFRA in LEO satellite communications.

\begin{algorithm}[t]
    \caption{EM-MRF-VAMP With BG Prior}
    \label{alg:Framwork}
    \begin{algorithmic}[1]
      \Require
        Observed signal $\mathbf{y}$, measurement matrix $\boldsymbol{\Gamma}$
         and the number of iterations $K_{\mathrm{it}}$ and $K_{\mathrm{mrf}}$.
      \State Initialization: $\forall i,j$: set $\mathbf{r}_{i,j}^{1,0}$ and $\gamma_{i,j}^{1,0}>0$, $\forall e,i,j,d:$ set $\pi_{i,j}^{e,d} = 0.5$,  $\alpha=\beta = 0.4$.
      \For {$k = 0 \to K_{\mathrm{it}}$}
      \State $\forall i,j: \widehat{\mathbf{g}}_{i,j}^{1k} 
             = \mathbf{f}_{1}\left(\mathbf{r}_{i,j}^{1k},\gamma_{i,j}^{1k}\right)$
             \vspace{3pt}
      \State $\forall i,j: \eta^{1k}_{i,j} = \left(\gamma^{1k}_{i,j} +(\phi_{i,j})^{-1} \right)/\left\langle \boldsymbol{\pi}_{i,j}^{k}\right\rangle$\vspace{3pt}
      \State $\forall i,j: \gamma_{i,j}^{2k} = \eta_{i,j}^{1k}-\gamma_{i,j}^{1k}$\vspace{3pt}
      \State $\forall i,j: \mathbf{r}_{i,j}^{2k} = \left(\eta_{i,j}^{1k}\widehat{\mathbf{g}}_{i,j}^{1k}-\gamma_{i,j}^{1k}\mathbf{r}_{i,j}^{1k}\right)/\gamma_{i,j}^{2k}$\vspace{2pt}   
      \State $\forall i,j: \widehat{\mathbf{g}}_{i,j}^{2k} 
             = \mathbf{f}_{2}\left(\mathbf{r}_{i,j}^{2k},\gamma_{i,j}^{2k}\right)$\vspace{2pt}
      \State $\forall i,j: \eta^{2k}_{i,j} = MN/\sum_{n}\frac{1}{\gamma^{2k}_{i,j}+|s_{n}|^{2}\sigma_{i,j}^{-2}}$\vspace{2pt}
      \State $\forall i,j: \gamma_{i,j}^{1,k+1} = \eta_{i,j}^{2k}-\gamma_{i,j}^{2k}$\vspace{2pt}
      \State $\forall i,j: \mathbf{r}_{i,j}^{1,k+1} = \left(\eta_{i,j}^{2k}\widehat{\mathbf{g}}^{2k}_{i,j}-\gamma_{i,j}^{2k}\mathbf{r}_{i,j}^{2k}\right)/\gamma_{i,j}^{1,k+1}$
      \State \%\textbf{MRF module}
      \State $\forall e,i,j$: Compute $\varpi_{q_{i,j}^{e}}^{s_{i,j}^{e}}$ via (\ref{xxqs})
  \For {$k_{\mathrm{mrf}} = 0 \to K_{\mathrm{mrf}}$}
      \State $\forall e,i,j$: Update $\pi_{i,j}^{e,t}$, $\pi_{i,j}^{e,b}$, $\pi_{i,j}^{e,l}$, $\pi_{i,j}^{e,r}$ via (27)
      \EndFor
      \State $\forall e,i,j$: Compute $\varpi^{q_{i,j}^{e}}_{s_{i,j}^{e}}$ via (\ref{xxsq})
      \State \%\textbf{Update hyperparameters}  
      \State $\forall i,j:$ Update $\sigma_{i,j}^{2}$, $\mu_{i,j}$, $\phi_{i,j}$ via (\ref{emsmp}).
      \State \textbf{if} $\sum_{i,j}\|\widehat{\mathbf{g}}^{1,k+1}_{i,j}-\widehat{\mathbf{g}}^{1k}_{i,j}\|_{F}^{2}<\tau\sum_{i,j}\|\widehat{\mathbf{g}}^{1k}_{i,j}\|_{F}^{2}$, \textbf{stop}
      \EndFor
      \State \textbf{Return:} Estimated parameters $\widehat{\mathbf{g}}^{1}$.
    \end{algorithmic}
  \end{algorithm}

  \begin{figure}[ht]
    \centering
\includegraphics[width=0.35\textwidth]{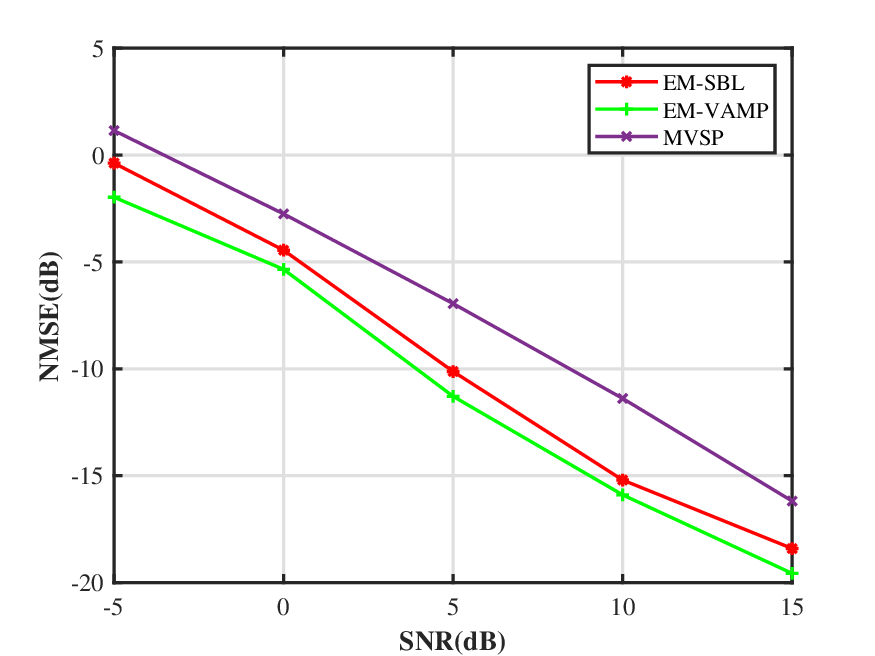}
    \caption{Performance comparison of channel estimation. $N_{y} = N_{z}=1$, $U = 40$, $p_{\lambda} = 0.2$, $M=2$, $N=64$, $f_{\mathrm{max}}$= 4 kHz.}
\end{figure}

\section{Numerical Results}
In this section, we conduct simulations to evaluate the performance of the proposed algorithm. According to 3GPP\cite{b13}, the scenarios about the non-terrestrial networks (NTN) are adopted. In our simulations, the satellite operates at the altitude of 600 km. The NTN-TDL-D channel is utilized, where the delay spread is 30 ns and the Doppler shift is uniformly distributed between [$-f_{\mathrm{max}}$, $f_{\mathrm{max}}$]. The sporadic transmission is considered, where the number of active devices depends on the number of devices $U$ and the active probability $p_{\lambda}$. The threshold $\xi$ is the same for each algorithms. In the end, the signal-to-noise ratio (SNR) is defined as SNR = $10\mathrm{log}_{10}\frac{\sum_{i}\sum_{j}\|\boldsymbol{\Gamma}\mathbf{g}_{i,j}\|_{F}^{2}}{\sigma^{2}NMN_{y}N_{z}}$. In order to measure the performance of channel estimation and device activity detection, the normalized mean-square-error (NMSE) and average device activity error rate (AER) are introduced, respectively. The two metrics are respectively given by NMSE $= \frac{\sum_{i}\sum_{j}\|\mathbf{H}^{A}-\hat{\mathbf{H}}^{A}\|_{F}^{2}}{\sum_{i}\sum_{j}\|\mathbf{H}^{A}\|_{F}^{2}}$ and AER $= \frac{1}{U} \sum_{u}{|\lambda_{u}-\widehat{\lambda}_{u}|}.$

Fig. 2 compares the channel estimation performance between our proposed algorithm and benchmarks, including the modified VSP (MVSP)\cite{b8} and EM-SBL\cite{sbl}. Our proposed EM-MRF-VAMP degenerates into EM-VAMP\cite{b12} since we set $N_{y}=N_{z}=1$. As mentioned above, the MVSP can only do well in the scenario with low Doppler. Therefore, we set $f_{\mathrm{max}}$ = 4 kHz, $\Delta f$ = 15 kHz and $\xi = 5$. As shown in Fig. 2, the performance of EM-VAMP increases with SNR and always outperforms MVSP. For example, when SNR = 5 dB, the NMSE of EM-VAMP is 4.3 dB better than MVSP. Additionally, though the EM-SBL achieves similar performance with EM-VAMP, its complexity is much higher than that of EM-VAMP.

\begin{figure}[ht]
    \centering
\includegraphics[width=0.35\textwidth]{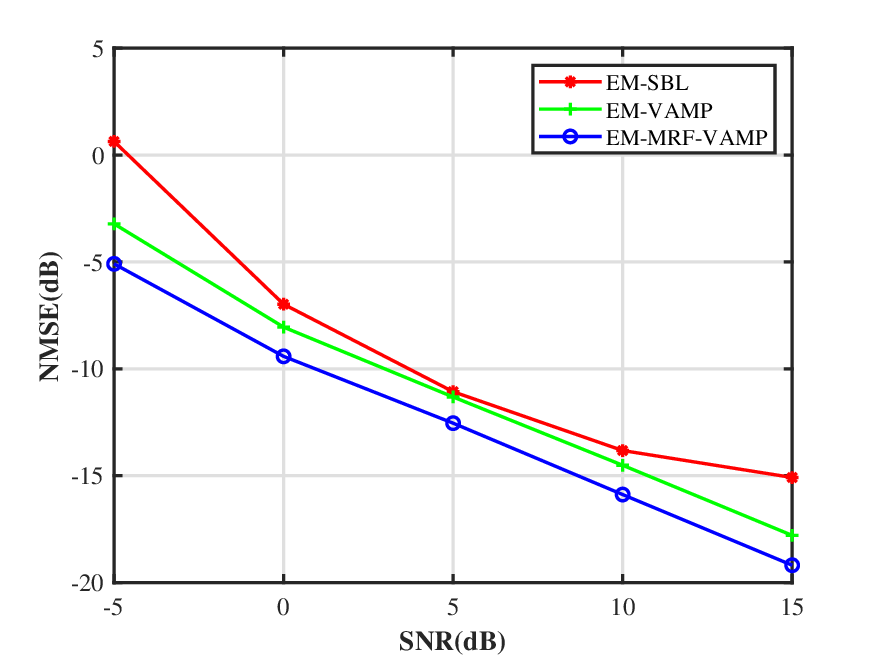}
    \caption{Performance comparison of channel estimation. $N_{y} = N_{z}=4$, $U = 100$, $p_{\lambda} = 0.1$, $M=8$, $N=32$, $f_{\mathrm{max}}$= 30 kHz.}
\end{figure}

Fig. 3 and Fig. 4 compare the channel estimation and device activity detection performance between our proposed EM-MRF-VAMP and benchmarks, respectively. We set $N_{y}=N_{z}=4$, $f_{\mathrm{max}}$ = 30 kHz, $\Delta f$ = 240 kHz and $\xi$ = 30. As shown in the figures, with the same SNR, the performance of the EM-MRF-VAMP outperforms the two benchmarks, indicating the superiority of our proposed algorithm. For example, when SNR = 10 dB in Fig.3, the NMSE of the proposed algorithm is 1.3 dB and 2 dB better than EM-VAMP and EM-SBL, respectively.

\begin{figure}[ht]
    \centering
\includegraphics[width=0.35\textwidth]{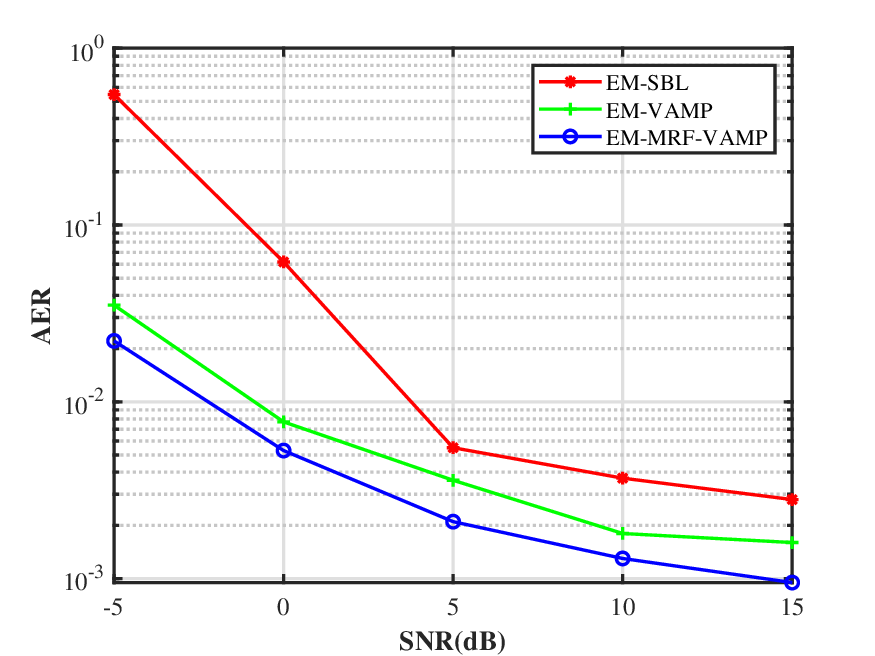}
    \caption{Performance comparison of device activity detection. $N_{y} = N_{z}=4$, $U = 100$, $p_{\lambda} = 0.1$, $M=8$, $N=32$, $f_{\mathrm{max}}$= 30 kHz.}
\end{figure}

\section{Conclusion}
This work investigated joint device activity detection and channel estimation for OFDM-based GFRA in uplink LEO satellite communications. To handle the severe Doppler shifts, a transmission scheme was proposed, where the DPS-BEM was utilized to accurately approximate the channel. Then, the VAMP algorithm was adopted to estimate the channel parameters, and the MRF module was combined to capture the channel sparsity. Finally, the hyperparameters were updated by the EM method. Simulation results demonstrated that the proposed transmission scheme and algorithm outperform traditional algorithms in terms of device activity detection and channel estimation.

\end{document}